\documentclass[conference]{IEEEtran}
\IEEEoverridecommandlockouts

\usepackage{cite}
\usepackage{amsmath,amssymb,amsfonts}
\usepackage{graphicx}
\usepackage{textcomp}
\usepackage{xcolor}
\usepackage{booktabs}
\usepackage{multirow}
\usepackage{orcidlink}
\usepackage{url}
\def\BibTeX{{\rm B\kern-.05em{\sc i\kern-.025em b}\kern-.08em
    T\kern-.1667em\lower.7ex\hbox{E}\kern-.125emX}}

\begin{document}

\title{Where Does Timbre Hide? Localizing Speaker Information in Speech Signals}

\author{Yiquan~Zhou~\orcidlink{0009-0002-1453-3820}, Yixuan~Wang~\orcidlink{0009-0002-3179-3894}, Xiaozhe~Qi~\orcidlink{0009-0006-9643-2116}, Xin~Gao~\orcidlink{0009-0007-4343-5345}, and Shihao~Li~\orcidlink{0009-0008-1989-2842}
\thanks{Yiquan Zhou, Yixuan Wang, and Xiaozhe Qi are with the School of Software Engineering, Xi'an Jiaotong University, Xi'an, China, and also with the SYKI-SPEECH Team, Xi'an, China (e-mail: yiqian.zhou@stu.xjtu.edu.cn; defender@stu.xjtu.edu.cn; qixz25@stu.xjtu.edu.cn).}
\thanks{Xin Gao and Shihao Li are with the Division of Music and Audio, Union Wheatland Culture and Media Ltd., Chengdu, China (e-mail: sui@musinya.com; yiseho@yiseho.com).}
\thanks{Corresponding author: Yiquan Zhou.}}

\maketitle

\begin{abstract}
Timbre similarity is a key evaluation target for generative speech systems, including voice conversion, speaker-cloning text-to-speech, and zero-shot speech models.
When the generated voice does not match the target speaker, the problem often reflects imperfect separation of speaker-related timbre from content, making it important to identify where timbre information resides and which representations still carry it.
Using speaker identity as a measurable proxy, we probe frequency bands, phase/magnitude structure, WORLD source/filter decomposition, voiced/unvoiced regions, self-supervised learning (SSL) layers, EnCodec/HuBERT tokens, and a 45-listener test across Mandarin, English, and Japanese speech/singing data.
Speaker information is distributed: 1--8 kHz magnitude and vocal-tract spectral-envelope structure dominate perception, above-8 kHz bands remain machine-discriminative, shallow SSL layers expose compact speaker information, and token leakage depends on tokenizer/layer choice.
These findings indicate where generation systems should condition on or transform speaker information, and where content, codec, and token pipelines should audit residual speaker leakage.
\end{abstract}


\section{Introduction}
\IEEEPARstart{T}{imbre} similarity is now a central evaluation target for generative speech systems.
Voice conversion (VC), speaker-cloning text-to-speech (TTS), and some zero-shot large speech models must preserve linguistic content while controlling the target speaker's voice~\cite{sisman2020overview,wang2025valle,zhang2024speechtokenizer,ju2024naturalspeech3}.
This requirement makes speaker-independent content modeling a practical engineering concern rather than only a design choice in representation learning.
If speaker-related information remains in a content, codec, or self-supervised learning (SSL) representation, it can leak into generation; if the information needed for the target speaker is not injected or transformed at the right place, the generated speech may remain intelligible but sound like the wrong speaker.

It is therefore important to know where speaker-related timbre information resides and how it is coupled with content and other acoustic factors.
Timbre itself is perceptual and difficult to measure directly at scale, so we use speaker identity as a measurable proxy for speaker-related timbre while keeping machine discriminability separate from human similarity judgments.
Classical source-filter theory and perceptual studies point to vocal-tract and spectral-envelope structure~\cite{fant1971acoustic,lavner2000effects,remez1997talker,kreiman2011foundations}, while modern speaker verification and SSL models such as ECAPA-TDNN and WavLM provide strong but less interpretable speaker information~\cite{desplanques2020ecapa,chen2022wavlm}.
Earlier sub-band speaker-recognition studies also found that low- and mid-frequency bands carry substantial speaker information~\cite{hayakawa1994text,besacier2000localization,kinnunen2010overview}, but they predate high-sample-rate corpora, neural codecs, tokenized speech language models, and recent SSL probing work~\cite{yang2021superb,pasad2021layer,shah2022what,defossez2023high}.

This paper asks where speaker-related timbre information is preserved, transformed, or leaked in both speech signals and learned representations.
We systematically compare complementary probes: frequency filtering, phase/magnitude randomization, WORLD source/filter swaps, voiced/unvoiced extraction, WavLM and ECAPA speaker verification, mel classifiers, SSL-layer and PCA probes, EnCodec and HuBERT token histograms, cross-style robustness checks, and a 45-listener similarity test.
These probes allow the same speaker-related information to be viewed from signal, representation, and perceptual sides.

The experiments lead to three findings.
First, the combined probes give a structured view of where speaker-related timbre is concentrated.
Frequency-band manipulations and listener ratings place the perceptually important speaker information mainly in the 1--8 kHz region, rather than in very high bands.
Phase/magnitude edits show that this information is carried mainly by spectral magnitude, not phase.
WORLD source/filter swaps further identify the vocal-tract spectral envelope as a stronger carrier than fundamental frequency (F0) alone.
Second, representation probes show that speaker-related information remains accessible in shallow SSL layers and can survive in some token streams, so content and codec representations should be audited rather than assumed speaker-free.
Third, speaker information measured by a classifier is not the same as perceived timbre similarity.
High-frequency bands still contain speaker information that classifiers can use, but this information does not produce strong same-speaker similarity for listeners.
This suggests that high-band reconstruction may improve acoustic detail without restoring the perceived speaker similarity that listeners rely on.
Together, these findings can guide future VC, TTS, anonymization, codec, and speech language model work in deciding which components need speaker conditioning, transformation, suppression, or leakage checks.

\section{Experimental Setup}
\label{sec:protocol}

This section defines the experimental setting used in the paper.
It contains two parts: the datasets and common evaluation setup, and the design of each experiment.

\subsection{Datasets and Common Setup}
The protocol uses multiple datasets because different probes require different data characteristics (e.g., sampling rate, language, style labels).
AISHELL-1 and AISHELL-3 provide Mandarin multi-speaker speech for frequency, SSL, token, and listening checks~\cite{bu2017aishell,shi2021aishell3}.
VCTK provides 48 kHz English speech for high-band and voiced/unvoiced tests~\cite{yamagishi2019cstr}.
ESD and CREMA-D test whether speaker information survives emotion changes~\cite{zhou2022esd,cao2014crema}.
NUS-48E and JVS/JVS-MuSiC test speech--singing transfer~\cite{duan2013nus,takamichi2019jvs,tamaru2020jvsmusic}.
We also use a self-recorded 48 kHz Chinese–English corpus for controlled band-swap and voiced/unvoiced pilot experiments.
Table~\ref{tab:scope} summarizes the data used by each experiment group.

\begin{table}[t]
\centering
\caption{Data used in the main checks.}
\label{tab:scope}
\footnotesize
\setlength{\tabcolsep}{2.2pt}
\begin{tabular}{p{0.23\columnwidth}p{0.16\columnwidth}p{0.50\columnwidth}}
\toprule
Experiment group & Data scale & What it checks \\
\midrule
AISHELL-1 & 20 spk. & lowpass/highpass, SSL, tokens \\
AISHELL-3 & 98 spk. & high-sample-rate bands, SSL/token validation, listening \\
VCTK & 100 spk. & 48 kHz high-band and unvoiced validation \\
ESD/CREMA-D & 20/91 spk. & emotion robustness \\
NUS/JVS & 12/100 spk. & speech--singing robustness \\
Self-rec. & 4 spk. & controlled band-swap and voiced/unvoiced pilot \\
\bottomrule
\end{tabular}
\end{table}

\begin{figure*}[t]
    \centering
    \includegraphics[width=0.92\textwidth]{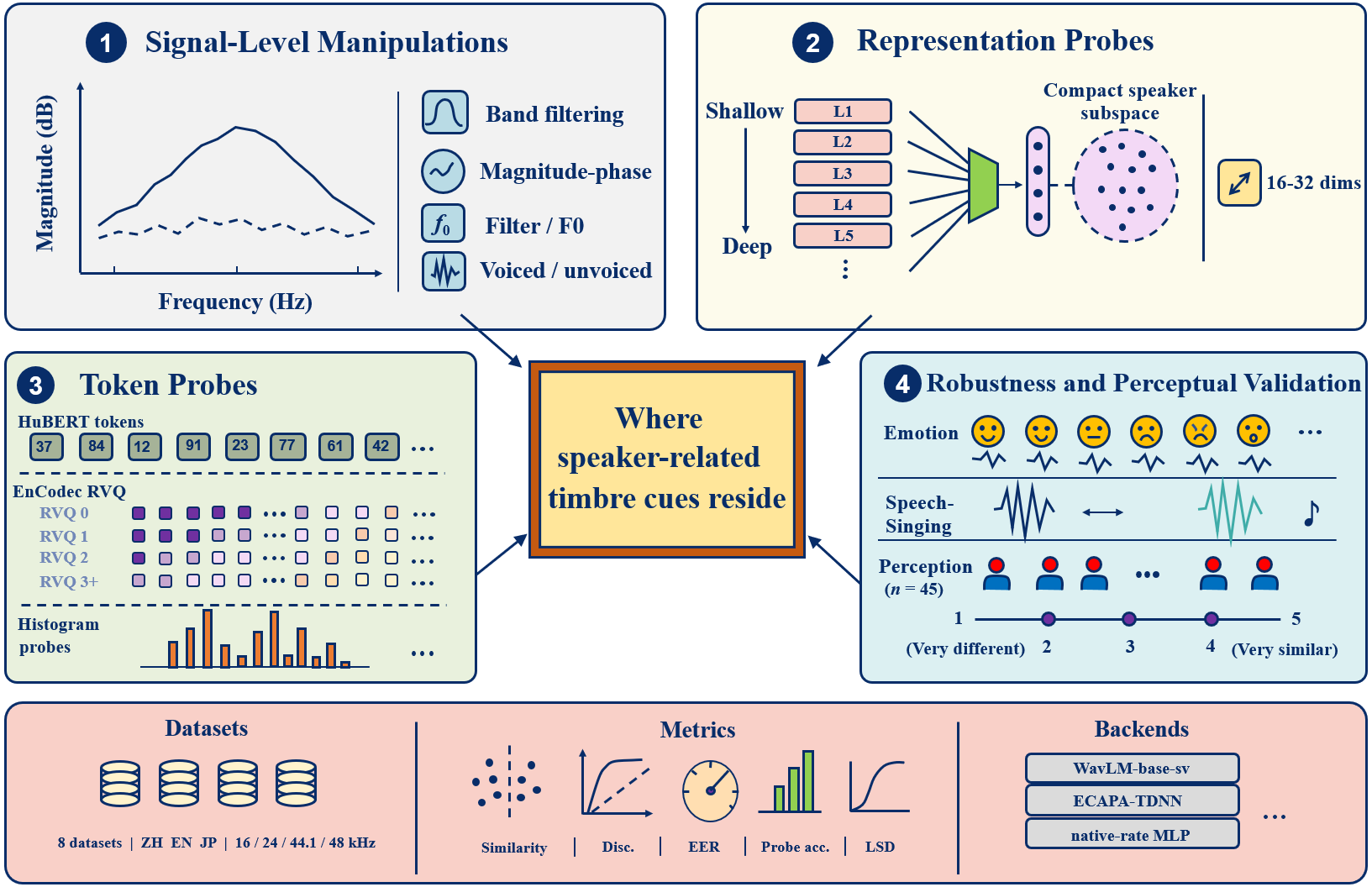}
    \caption{Speaker-information localization protocol. The study combines signal edits, model-based and model-free measurements, representation probes, and human similarity ratings, so speaker information measured by models can be kept separate from perceived speaker-related timbre information.}
    \label{fig:protocol}
\end{figure*}

All experiments use speaker identity as the measurable target, but the metric depends on the probe.
WavLM-base-sv is the main speaker verification (SV) backend~\cite{chen2022wavlm}.
For signal edits that may depend on the backend, we repeat the phase/magnitude and source/filter tests with a SpeechBrain ECAPA-TDNN VoxCeleb model~\cite{desplanques2020ecapa}.
Embedding-based tests use cosine similarity between speaker embeddings, and report intra-speaker similarity, inter-speaker similarity, and discriminability, defined as the same-speaker score minus the different-speaker score.
Bandwidth curves also report equal error rate (EER).
Closed-set feature and token probes report speaker-classification accuracy.
The listening test reports mean voice-timbre similarity ratings (i.e., perceived same-speaker similarity) on a 1--5 scale.
Because WavLM-base-sv requires 16 kHz input, spectral content above 8 kHz is discarded before embedding extraction.
High-sample-rate spectra are therefore evaluated with mel multilayer perceptron (MLP) probes in the frequency-band experiment.
Fig.~\ref{fig:protocol} summarizes the full protocol.

\subsection{Experiment Design}
\textbf{Frequency-band experiment:}
This experiment tests which spectral regions preserve speaker information.
For lowpass and highpass analysis, we apply Butterworth filters to each utterance, extract WavLM-base-sv embeddings, and compute speaker discriminability and EER.
Lowpass curves show how speaker information changes as more bandwidth is retained from the bottom up.
Highpass curves show how speaker information changes when low and mid frequencies are removed.
For bands above 8 kHz, we cannot rely on WavLM-base-sv because its 16 kHz input removes this frequency range.
We therefore keep AISHELL-3 and VCTK at their original sampling rates, isolate fixed frequency bands, extract log-mel summaries from each band, and train a 2-layer closed-set MLP speaker classifier.
We also check shuffled labels and duration/energy baselines to make sure the classifier is not succeeding for trivial reasons.
Mel-cosine separability (the cosine-distance gap between same- and different-speaker mel features) provides an additional classifier-free check.
To compare machine-detectable information with perception, we also construct AISHELL-3 paired trials from full-band, lowpass, bandpass, highpass, and different-speaker conditions.
Listeners rate each pair on a 1--5 voice-timbre similarity scale.
The main metrics are discriminability and EER for the SV curves, speaker-classification accuracy for the mel MLP checks, and mean listener rating for the listening conditions.

\textbf{Phase/magnitude experiment:}
This experiment tests whether speaker information is carried mainly by spectral magnitude or phase.
For each utterance, we compute a short-time Fourier transform (STFT) and synthesize two variants.
The magnitude-kept variant keeps the original magnitude and randomizes phase.
The phase-kept variant keeps the original phase and randomizes magnitude.
We then compare the speaker discriminability of each variant against that of the unmodified signal.
This manipulation directly tests whether speaker information is carried by the spectral envelope or the phase structure.

\textbf{Source/filter experiment:}
This experiment tests whether speaker information is better explained by source pitch (F0) or vocal-tract filtering.
Using WORLD, we decompose each utterance into fundamental frequency (F0), spectral envelope, and aperiodicity~\cite{morise2016world}.
We then create cross-speaker resynthesized signals by keeping one speaker's F0 while replacing the spectral envelope, or by keeping the spectral envelope while replacing F0.
We judge the two conditions by same-speaker cosine similarity and retention (the fraction of the original same-speaker similarity preserved).
If the filter-kept signal preserves more speaker similarity than the F0-kept signal, the retained information is more closely linked to vocal-tract spectral-envelope structure than to pitch alone.
We repeat the phase/magnitude and source/filter tests with ECAPA-TDNN to check that the pattern is not specific to WavLM-base-sv.

\textbf{Self-supervised learning (SSL) representation experiment:}
This experiment tests where speaker information is exposed in continuous speech features.
We extract layer-wise features from WavLM, HuBERT, Wav2Vec 2.0, and Whisper-base, mean-pool each utterance, and train lightweight speaker classifiers.
We also apply principal component analysis (PCA) before classification to test whether speaker information remains accessible after dimensionality reduction.
This shows which layers are most speaker-discriminative and whether the information is compact enough to leak through content features when they are reused in downstream systems.
The main metric is closed-set speaker-classification accuracy.

\textbf{Token experiment:}
This experiment tests whether discrete speech units retain speaker information.
For EnCodec, we compute histograms from each residual vector quantization (RVQ) layer and from cumulative RVQ groups.
For HuBERT, we compute k-means token histograms from multiple SSL layers and codebook sizes, and add a bigram check to capture limited sequential dependencies.
These probes test whether speaker leakage depends on tokenizer, layer, codebook size, and RVQ depth.
The main metric is closed-set speaker-classification accuracy.

\textbf{Robustness experiments:}
These experiments test whether speaker information survives common style and phonation changes.
Emotion tests compare same-speaker cross-emotion similarity with different-speaker baselines, and include a CREMA-D leave-one-emotion-out validation.
Speech--singing tests compare same-speaker cross-modal similarity and train speech-based classifiers that are evaluated on singing.
Voiced/unvoiced tests separate active speech into voiced and unvoiced regions and evaluate whether speaker information survives outside pitch-bearing regions.
These checks prevent the main claims from relying only on neutral read speech.
The main metrics are cosine similarity gaps and speaker-classification accuracy.

\section{Results}
\label{sec:results}

\subsection{Frequency-Band Experiment Results}
The frequency-band results show that the strongest speaker information lies in the low/mid spectrum, while higher bands retain weaker but measurable information.
In the WavLM-base-sv lowpass curves, discriminability increases as more bandwidth is retained and nearly reaches the full-band score by 7.5 kHz.
The highpass curves give the complementary pattern: performance remains stable when only sub-1 kHz content is removed, but drops sharply once the cutoff exceeds 1--2 kHz.
These two curves indicate that speaker information is not carried by sub-1 kHz content alone, but by a broader low/mid-frequency range.
Because WavLM-base-sv is limited by its 16 kHz input, the WavLM curves cannot determine whether bands above 8 kHz still contain speaker information.

\begin{table}[t]
\centering
\caption{Band MLP check. Values are speaker-classification accuracy (\%); shuffled labels stay near chance and duration/energy baselines are much lower.}
\label{tab:native_band_mlp}
\footnotesize
\setlength{\tabcolsep}{2.4pt}
\begin{tabular}{lcccc}
\toprule
Dataset/Band & 1--4 & 4--8 & 8--12 & High band \\
\midrule
VCTK acc. & 92.1 & 89.3 & 88.1 & 50.1 (20--24) \\
VCTK shuf. & 1.2 & 1.5 & 1.1 & 0.8 \\
VCTK dur./eng. & 4.5 & 5.8 & 5.4 & 5.4 \\
\midrule
AISHELL-3 acc. & 82.5 & 72.4 & 55.0 & 42.8 (16--20) \\
AISHELL-3 shuf. & 0.9 & 1.0 & 1.0 & 0.9 \\
AISHELL-3 dur./eng. & 6.4 & 5.6 & 5.3 & 6.0 \\
\bottomrule
\end{tabular}
\end{table}

\begin{figure}[t]
    \centering
    \includegraphics[width=\columnwidth]{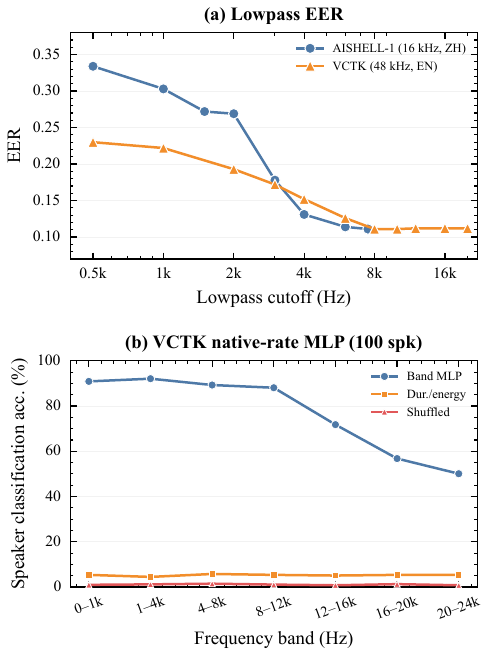}
    \caption{Bandwidth comparison. SV lowpass curves saturate near 4--8 kHz, while mel probes show weaker but measurable above-8 kHz information.}
    \label{fig:bandwidth}
\end{figure}

Table~\ref{tab:native_band_mlp} fills this gap with the mel MLP checks.
Both high-sample-rate datasets show the same trend: accuracy decreases with frequency.
The 1--8 kHz bands are strongest: VCTK reaches 92.1\% at 1--4 kHz and 89.3\% at 4--8 kHz, while AISHELL-3 reaches 82.5\% and 72.4\% in the same bands.
Bands above 8 kHz remain well above shuffled labels and duration/energy baselines, so they do contain speaker information that classifiers can use.
However, the scores decline with frequency: VCTK falls to 50.1\% in the 20--24 kHz band, and AISHELL-3 falls to 42.8\% in the 16--20 kHz band.
Mel-cosine separability confirms the same frequency trend, with AISHELL-3 spectral-shape features peaking at 4--8 kHz ($d'=1.86$) and decreasing toward the highest tested band ($d'=0.97$).
Fig.~\ref{fig:bandwidth} summarizes the contrast between low/mid-band dominance and weaker high-band information.

\begin{figure}[t]
    \centering
    \includegraphics[width=\columnwidth]{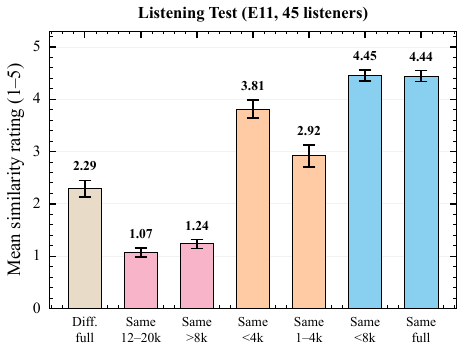}
    \caption{Listening test. Low/mid-frequency same-speaker pairs receive high similarity ratings; high-frequency-only pairs receive low ratings.}
    \label{fig:listening}
\end{figure}

Table~\ref{tab:listening} and Fig.~\ref{fig:listening} report the listening-test results for the same frequency contrast.

\begin{table}[t]
\centering
\caption{Human similarity ratings (45 listeners, 1--5 scale).}
\label{tab:listening}
\footnotesize
\setlength{\tabcolsep}{3pt}
\begin{tabular}{lcc}
\toprule
Condition & Mean $\pm$ SD & 95\% CI \\
\midrule
Different full & 2.29 $\pm$ 0.53 & [2.13, 2.45] \\
Same 12--20 kHz & 1.07 $\pm$ 0.28 & [0.99, 1.16] \\
Same $>$8 kHz & 1.24 $\pm$ 0.29 & [1.15, 1.32] \\
Same $<$4 kHz & 3.81 $\pm$ 0.58 & [3.64, 3.99] \\
Same 1--4 kHz & 2.92 $\pm$ 0.69 & [2.71, 3.13] \\
Same $<$8 kHz & \textbf{4.45 $\pm$ 0.35} & [4.35, 4.56] \\
Same full & 4.44 $\pm$ 0.36 & [4.34, 4.55] \\
\bottomrule
\end{tabular}
\end{table}

The listening results show that classifier-usable high-frequency information is not sufficient for perceived same-speaker timbre similarity.
Same-speaker full-band pairs are rated much higher than different-speaker full-band controls (4.44 vs. 2.29, $p=3.6\times10^{-21}$).
Low/mid-frequency same-speaker conditions (Same $<$4 kHz, Same 1--4 kHz, Same $<$8 kHz) are rated much higher than high-frequency-only conditions (Same $>$8 kHz, Same 12--20 kHz), with a mean paired difference of 2.57 points ($p=1.7\times10^{-38}$).
The $<8$ kHz versus $>8$ kHz contrast is especially large (4.45 vs. 1.24, $p=1.3\times10^{-39}$).
Together with the MLP results, this shows that high-frequency bands contain detectable speaker information, but do not provide strong same-speaker similarity for listeners under the present manipulations.

\subsection{Phase/Magnitude Experiment Results}
\begin{figure}[t]
    \centering
    \includegraphics[width=\columnwidth]{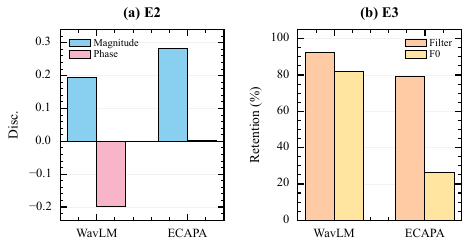}
    \caption{Second-backend validation. WavLM and ECAPA both favor magnitude/filter information over phase/F0-dominant information.}
    \label{fig:e2e3}
\end{figure}

Table~\ref{tab:e2e3} and Fig.~\ref{fig:e2e3} summarize the phase/magnitude and source/filter results across the two SV backends.
For the phase/magnitude experiment, magnitude preserves much more speaker information than phase.
With WavLM, keeping the original magnitude spectrum while randomizing phase preserves 81.6\% of the original discriminability.
The phase-kept variant loses speaker discriminability entirely: WavLM discriminability drops to -0.199 and ECAPA to 0.003, both effectively at or below chance.

\begin{table}[t]
\centering
\caption{Phase/magnitude and source/filter summary across two SV backends. Ret. denotes retention relative to the full or original condition.}
\label{tab:e2e3}
\footnotesize
\setlength{\tabcolsep}{2.2pt}
\begin{tabular}{llcc}
\toprule
Backend & Condition & Score & Ret. \\
\midrule
WavLM & magnitude kept & disc. 0.195 & 81.6\% \\
WavLM & phase kept & disc. -0.199 & -- \\
WavLM & filter kept & sim. 0.906 & 92.4\% \\
WavLM & F0 kept & sim. 0.804 & 82.0\% \\
\midrule
ECAPA & magnitude kept & disc. 0.284 & 77.1\% \\
ECAPA & phase kept & disc. 0.003 & 0.7\% \\
ECAPA & filter kept & sim. 0.601 & 79.3\% \\
ECAPA & F0 kept & sim. 0.199 & 26.2\% \\
\bottomrule
\end{tabular}
\end{table}

\subsection{Source/Filter Experiment Results}
The source/filter rows in Table~\ref{tab:e2e3} and Fig.~\ref{fig:e2e3} confirm the same pattern in a source-filter decomposition: spectral-envelope information dominates over F0.
Keeping the vocal-tract spectral envelope preserves 92.4\% of full-signal same-speaker similarity with WavLM, compared with 82.0\% when only F0 is retained.
With ECAPA, the filter-kept condition also retains much more speaker similarity than the F0-kept condition (79.3\% vs. 26.2\%).
Together with the phase/magnitude result in Table~\ref{tab:e2e3}, this links the main speaker-related information to spectral-envelope structure rather than to phase or F0 alone.

\subsection{SSL Representation Experiment Results}
Table~\ref{tab:ssl_checks} summarizes the model-wise layer and PCA results.
The representation results show that speaker information is strongest in early to mid layers.
On AISHELL-1, WavLM, HuBERT, Wav2Vec 2.0, and Whisper speaker probes all peak in the first few layers (L1--L5), well before the final layer.
On AISHELL-3 (not tabulated), WavLM reaches 96.5--97.7\% in layers L0--L4 but falls to 75.7--76.5\% in L9--L12.
PCA results show the same depth effect: AISHELL-3 WavLM L0 remains at 95.8\% with 32 dimensions, while L12 falls to 62.7\%.

\begin{table}[t]
\centering
\caption{SSL representation results on AISHELL-1. PCA columns use each model's peak layer.}
\label{tab:ssl_checks}
\begingroup
\footnotesize
\setlength{\tabcolsep}{1.9pt}
\renewcommand{\arraystretch}{1.08}
\begin{tabular}{@{}lccccc@{}}
\toprule
Model & Peak & Final & 32d & 16d & 8d \\
\midrule
WavLM & L1 99.8 & L12 94.8 & 99.8 & 98.3 & 85.3 \\
HuBERT & L5 99.5 & L12 95.3 & 99.0 & 93.8 & 69.2 \\
Wav2Vec2 & L1 99.3 & L12 95.0 & 99.0 & 95.8 & 82.2 \\
Whisper & L3 97.5 & L6 92.0 & 96.5 & 90.2 & 74.5 \\
\bottomrule
\end{tabular}
\endgroup
\end{table}

\subsection{Token Experiment Results}
Table~\ref{tab:token_checks} summarizes the token and codec results.
The results are tokenizer- and layer-dependent.
With the order-free histogram probe used here, HuBERT k-means tokens preserve more speaker information than the tested 24 kHz EnCodec RVQ histograms.
This is not a final ranking of all codec representations.
These results show that tokenization objective and layer choice strongly affect how much speaker information remains available.

\begin{table}[t]
\centering
\caption{Token and codec results. HuBERT and EnCodec values are speaker-classification accuracies (\%).}
\label{tab:token_checks}
\footnotesize
\setlength{\tabcolsep}{1.9pt}
\renewcommand{\arraystretch}{1.08}
\begin{tabular}{@{}lccccc@{}}
\toprule
\multicolumn{6}{c}{\textbf{HuBERT k-means tokens}} \\
\midrule
Layer & $k$=50 & $k$=100 & $k$=200 & $k$=500 & $k$=1000 \\
\midrule
L1  & 60.2 & 68.5 & 78.0 & 89.0 & 94.0 \\
L3  & 46.5 & 65.7 & 70.7 & 84.0 & 92.7 \\
L6  & 43.0 & 57.8 & 75.3 & 87.3 & \textbf{96.2} \\
L9  & 37.7 & 52.0 & 64.7 & 75.2 & 88.5 \\
L12 & 40.7 & 57.2 & 68.3 & 74.2 & 84.2 \\
\midrule
\multicolumn{6}{c}{\textbf{EnCodec RVQ tokens}} \\
\midrule
Layer & \multicolumn{2}{c}{Acc.} & Cum. & \multicolumn{2}{c}{Cum.\ Acc.} \\
\midrule
RVQ-0 & \multicolumn{2}{c}{53.5} & 0    & \multicolumn{2}{c}{53.5} \\
RVQ-1 & \multicolumn{2}{c}{65.0} & 0--1 & \multicolumn{2}{c}{63.7} \\
RVQ-2 & \multicolumn{2}{c}{67.2} & 0--2 & \multicolumn{2}{c}{\textbf{68.8}} \\
RVQ-3 & \multicolumn{2}{c}{47.0} & 0--3 & \multicolumn{2}{c}{68.7} \\
RVQ-4 & \multicolumn{2}{c}{45.8} & 0--4 & \multicolumn{2}{c}{67.8} \\
RVQ-5 & \multicolumn{2}{c}{42.0} & 0--5 & \multicolumn{2}{c}{68.8} \\
RVQ-6 & \multicolumn{2}{c}{36.0} & 0--6 & \multicolumn{2}{c}{67.8} \\
RVQ-7 & \multicolumn{2}{c}{36.7} & 0--7 & \multicolumn{2}{c}{67.0} \\
\bottomrule
\end{tabular}
\end{table}

\subsection{Robustness Experiment Results}
\label{sec:robustness}
Table~\ref{tab:robust_details} reports the robustness results across style, modality, and phonation changes.
Speaker information remains detectable in all tested settings.
In CREMA-D, leave-one-emotion-out speaker classification over 91 actors reaches 89.2\% mean accuracy; anger is hardest at 75.4\%, still far above the 1.1\% chance level.
In JVS/JVS-MuSiC, a classifier trained on 10,000 read-speech utterances from 100 speakers reaches 23.7\% on 600 singing segments — above the 1.0\% chance level but substantially lower than within-mode accuracy.
For VCTK voiced/unvoiced validation, unvoiced-only active frames reach 70.9\% speaker accuracy, and the same-speaker cosine similarity gap between voiced and unvoiced frames is 0.463 (voiced higher).
These checks show that speaker information remains detectable across emotion, singing, and unvoiced speech, although it weakens when style or phonation changes.

\begin{table}[t]
\centering
\caption{Robustness details across style, modality, and phonation changes.}
\label{tab:robust_details}
\footnotesize
\setlength{\tabcolsep}{2.1pt}
\begin{tabular}{p{0.32\columnwidth}p{0.57\columnwidth}}
\toprule
Condition & Quantitative detail \\
\midrule
ESD emotion & Same-speaker cross-emotion cosine 0.899 vs. different-speaker same-emotion 0.644; gap +0.255. \\
CREMA-D & Leave-one-emotion-out mean 89.2\%; held-out anger 75.4\%, neutral 95.1\%. \\
NUS-48E & Sing--read same-speaker similarity 0.909; cross-modal discriminability 0.246. \\
JVS/JVS-MuSiC & Speech-trained classifier reaches 23.7\% on singing; centroid gap 0.272. \\
VCTK V/UV & Unvoiced-only 70.9$\pm$1.2\%; voiced-vs.-unvoiced cosine gap 0.463. \\
\bottomrule
\end{tabular}
\end{table}

\section{Discussion}
\label{sec:discussion}

The results provide a consistent localization pattern.
The strongest perceived speaker-related timbre information is carried by low- and mid-frequency magnitude structure, especially in the 1--8 kHz region.
Phase/magnitude edits and WORLD source/filter swaps, validated with a second SV backend (ECAPA-TDNN), further point to the vocal-tract spectral envelope rather than phase or F0 alone.
This agrees with the source-filter view of speech: the vocal tract shapes formants and broad spectral balance, both of which remain relatively stable across utterances; vocal-tract filtering also applies to unvoiced segments, consistent with the unvoiced speaker-recognition results in Section~\ref{sec:robustness}.

The high-frequency results need a separate interpretation.
Bands above 8 kHz remain useful for closed-set classifiers and mel-cosine checks, so they should not be treated as empty.
However, high-frequency-only same-speaker pairs receive low listener ratings.
This suggests that high-band reconstruction may contribute to acoustic detail (e.g., frication or airiness), but is unlikely to restore the speaker similarity that listeners primarily perceive.

Learned representations echo the signal-level findings.
Shallow SSL layers retain more speaker information and keep it in a compact subspace, while deeper layers expose less speaker information under our probes.
HuBERT and EnCodec results also show that discrete representations are not automatically speaker-free.
The amount of speaker information depends on the source layer, codebook size, and residual vector quantization layer.
Therefore, content, codec, and token pipelines should check speaker leakage at the actual representation used by the downstream system, not only in explicit speaker embeddings.

The robustness experiments show that this information is not limited to clean neutral voiced speech.
Speaker information remains measurable across emotion, speech--singing changes, and unvoiced segments.
The strength decreases under larger domain changes, especially speech-to-singing transfer, but remains above chance.
Combined with the signal-level findings, this supports the view that speaker-related spectral-envelope structure can survive changes in prosody, style, and phonation.

These findings provide a practical map for speaker-aware speech systems.
Voice conversion and speaker-cloning TTS should control low/mid spectral-envelope information when matching a target speaker, while treating high-band enhancement mainly as acoustic detail unless the downstream task specifically requires closed-set speaker discrimination at high sampling rates.
Anonymization and tokenized speech language models should test residual speaker information in shallow SSL features, codec layers, and token streams.
A single speaker-similarity score is therefore not enough: evaluation should separate machine speaker discriminability, perceived speaker similarity, and representation-level leakage.

\section{Limitations and Conclusion}
This study has several limitations.
Many probes are closed-set, so speaker-classification accuracy and SV-style scores should not be interpreted as open-set speaker verification performance.
WavLM-base-sv operates at 16 kHz, which limits SV-based frequency analysis above 8 kHz; the high-sample-rate mel probes, mel-cosine checks, and listening test reduce this issue but do not replace a high-sample-rate open-set SV backend.
The high-frequency listening stimuli are narrowband and less natural than full-band speech, so their low ratings describe perceived voice-timbre similarity under the present filtering conditions.
The emotion, singing, and token results would also benefit from more speaker backends, more detailed listener metadata, better-controlled playback conditions, and sequence-aware token probes.

Overall, this paper uses speaker identity as a measurable proxy to study where speaker-related timbre information resides in speech signals and learned representations.
First, the strongest perceived speaker-related information is concentrated in low/mid-frequency spectral-magnitude structure — specifically the vocal-tract spectral envelope in the 1--8 kHz region — while phase, F0, and high-frequency-only information is weaker under our probes.
Second, speaker information is not confined to explicit speaker embeddings: shallow SSL layers and some token streams still expose recoverable speaker information.
Third, machine speaker discriminability is not the same as perceived timbre similarity: high-frequency bands contain speaker information that classifiers can use, but they do not produce strong same-speaker ratings in the listening test.
Together, these findings indicate where VC, speaker-cloning TTS, anonymization, codec, and speech language model pipelines should inject, transform, suppress, or audit speaker-related information.

\clearpage
\bibliographystyle{IEEEtran}
\bibliography{ref}

\begin{thebibliography}{10}
\providecommand{\url}[1]{#1}
\csname url@samestyle\endcsname
\providecommand{\newblock}{\relax}
\providecommand{\bibinfo}[2]{#2}
\providecommand{\BIBentrySTDinterwordspacing}{\spaceskip=0pt\relax}
\providecommand{\BIBentryALTinterwordstretchfactor}{4}
\providecommand{\BIBentryALTinterwordspacing}{\spaceskip=\fontdimen2\font plus
\BIBentryALTinterwordstretchfactor\fontdimen3\font minus \fontdimen4\font\relax}
\providecommand{\BIBforeignlanguage}[2]{{%
\expandafter\ifx\csname l@#1\endcsname\relax
\typeout{** WARNING: IEEEtran.bst: No hyphenation pattern has been}%
\typeout{** loaded for the language `#1'. Using the pattern for}%
\typeout{** the default language instead.}%
\else
\language=\csname l@#1\endcsname
\fi
#2}}
\providecommand{\BIBdecl}{\relax}
\BIBdecl

\bibitem{sisman2020overview}
B.~Sisman, J.~Yamagishi, S.~King, and H.~Li, ``An overview of voice conversion and its challenges: From statistical modeling to deep learning,'' \emph{IEEE/ACM Transactions on Audio, Speech, and Language Processing}, vol.~29, pp. 132--157, 2021.

\bibitem{wang2025valle}
S.~Chen, C.~Wang, Y.~Wu, Z.~Zhang, L.~Zhou, S.~Liu, Z.~Chen, Y.~Liu, H.~Wang, J.~Li, L.~He, S.~Zhao, and F.~Wei, ``Neural codec language models are zero-shot text to speech synthesizers,'' \emph{IEEE Transactions on Audio, Speech and Language Processing}, vol.~33, pp. 705--718, 2025.

\bibitem{zhang2024speechtokenizer}
\BIBentryALTinterwordspacing
X.~Zhang, D.~Zhang, S.~Li, Y.~Zhou, and X.~Qiu, ``Speechtokenizer: Unified speech tokenizer for speech language models,'' in \emph{International Conference on Learning Representations}, B.~Kim, Y.~Yue, S.~Chaudhuri, K.~Fragkiadaki, M.~Khan, and Y.~Sun, Eds., vol. 2024, 2024, pp. 31\,798--31\,818. [Online]. Available: \url{https://proceedings.iclr.cc/paper_files/paper/2024/file/86d1ab582afb247ccaa84bec4a7e24f7-Paper-Conference.pdf}
\BIBentrySTDinterwordspacing

\bibitem{ju2024naturalspeech3}
\BIBentryALTinterwordspacing
Z.~Ju, Y.~Wang, K.~Shen, X.~Tan, D.~Xin, D.~Yang, E.~Liu, Y.~Leng, K.~Song, S.~Tang, Z.~Wu, T.~Qin, X.~Li, W.~Ye, S.~Zhang, J.~Bian, L.~He, J.~Li, and S.~Zhao, ``{N}atural{S}peech 3: Zero-shot speech synthesis with factorized codec and diffusion models,'' in \emph{Proceedings of the 41st International Conference on Machine Learning}, ser. Proceedings of Machine Learning Research, R.~Salakhutdinov, Z.~Kolter, K.~Heller, A.~Weller, N.~Oliver, J.~Scarlett, and F.~Berkenkamp, Eds., vol. 235.\hskip 1em plus 0.5em minus 0.4em\relax PMLR, 21--27 Jul 2024, pp. 22\,605--22\,623. [Online]. Available: \url{https://proceedings.mlr.press/v235/ju24b.html}
\BIBentrySTDinterwordspacing

\bibitem{fant1971acoustic}
\BIBentryALTinterwordspacing
G.~Fant, \emph{Acoustic Theory of Speech Production: With Calculations based on X-Ray Studies of Russian Articulations}.\hskip 1em plus 0.5em minus 0.4em\relax Berlin, Boston: De Gruyter Mouton, 1971. [Online]. Available: \url{https://doi.org/10.1515/9783110873429}
\BIBentrySTDinterwordspacing

\bibitem{lavner2000effects}
\BIBentryALTinterwordspacing
Y.~Lavner, I.~Gath, and J.~Rosenhouse, ``The effects of acoustic modifications on the identification of familiar voices speaking isolated vowels,'' \emph{Speech Communication}, vol.~30, no.~1, pp. 9--26, 2000. [Online]. Available: \url{https://www.sciencedirect.com/science/article/pii/S016763939900028X}
\BIBentrySTDinterwordspacing

\bibitem{remez1997talker}
R.~E. Remez, J.~M. Fellowes, and P.~E. Rubin, ``Talker identification based on phonetic information,'' \emph{Journal of Experimental Psychology: Human Perception and Performance}, vol.~23, no.~3, pp. 651--666, 1997.

\bibitem{kreiman2011foundations}
J.~Kreiman and D.~Sidtis, \emph{Foundations of voice studies: An interdisciplinary approach to voice production and perception}.\hskip 1em plus 0.5em minus 0.4em\relax Wiley-Blackwell, 2011.

\bibitem{desplanques2020ecapa}
B.~Desplanques, J.~Thienpondt, and K.~Demuynck, ``{ECAPA-TDNN: Emphasized Channel Attention, Propagation and Aggregation in TDNN Based Speaker Verification},'' in \emph{{Interspeech 2020}}, 2020, pp. 3830--3834.

\bibitem{chen2022wavlm}
S.~Chen, C.~Wang, Z.~Chen, Y.~Wu, S.~Liu, Z.~Chen, J.~Li, N.~Kanda, T.~Yoshioka, X.~Xiao, J.~Wu, L.~Zhou, S.~Ren, Y.~Qian, Y.~Qian, J.~Wu, M.~Zeng, X.~Yu, and F.~Wei, ``Wavlm: Large-scale self-supervised pre-training for full stack speech processing,'' \emph{IEEE Journal of Selected Topics in Signal Processing}, vol.~16, no.~6, pp. 1505--1518, 2022.

\bibitem{hayakawa1994text}
S.~Hayakawa and F.~Itakura, ``Text-dependent speaker recognition using the information in the higher frequency band,'' in \emph{Proceedings of ICASSP '94. IEEE International Conference on Acoustics, Speech and Signal Processing}, vol.~i, 1994, pp. I/137--I/140 vol.1.

\bibitem{besacier2000localization}
\BIBentryALTinterwordspacing
L.~Besacier, J.~Bonastre, and C.~Fredouille, ``Localization and selection of speaker-specific information with statistical modeling,'' \emph{Speech Communication}, vol.~31, no. 2-3, pp. 89--106, 2000. [Online]. Available: \url{https://www.sciencedirect.com/science/article/pii/S0167639399000709}
\BIBentrySTDinterwordspacing

\bibitem{kinnunen2010overview}
\BIBentryALTinterwordspacing
T.~Kinnunen and H.~Li, ``An overview of text-independent speaker recognition: From features to supervectors,'' \emph{Speech Communication}, vol.~52, no.~1, pp. 12--40, 2010. [Online]. Available: \url{https://www.sciencedirect.com/science/article/pii/S0167639309001289}
\BIBentrySTDinterwordspacing

\bibitem{yang2021superb}
S.~wen Yang, P.-H. Chi, Y.-S. Chuang, C.-I.~J. Lai, K.~Lakhotia, Y.~Y. Lin, A.~T. Liu, J.~Shi, X.~Chang, G.-T. Lin, T.-H. Huang, W.-C. Tseng, K.~tik Lee, D.-R. Liu, Z.~Huang, S.~Dong, S.-W. Li, S.~Watanabe, A.~Mohamed, and H.~yi~Lee, ``{SUPERB: Speech Processing Universal PERformance Benchmark},'' in \emph{{Interspeech 2021}}, 2021, pp. 1194--1198.

\bibitem{pasad2021layer}
A.~Pasad, J.-C. Chou, and K.~Livescu, ``Layer-wise analysis of a self-supervised speech representation model,'' in \emph{2021 IEEE Automatic Speech Recognition and Understanding Workshop (ASRU)}, 2021, pp. 914--921.

\bibitem{shah2022what}
Y.~K. Singla, J.~Shah, C.~Chen, and R.~R. Shah, ``What do audio transformers hear? probing their representations for language delivery \& structure,'' in \emph{2022 IEEE International Conference on Data Mining Workshops (ICDMW)}, 2022, pp. 910--925.

\bibitem{defossez2023high}
\BIBentryALTinterwordspacing
A.~D{\'e}fossez, J.~Copet, G.~Synnaeve, and Y.~Adi, ``High fidelity neural audio compression,'' \emph{Transactions on Machine Learning Research}, 2023, featured Certification, Reproducibility Certification. [Online]. Available: \url{https://openreview.net/forum?id=ivCd8z8zR2}
\BIBentrySTDinterwordspacing

\bibitem{bu2017aishell}
H.~Bu, J.~Du, X.~Na, B.~Wu, and H.~Zheng, ``Aishell-1: An open-source mandarin speech corpus and a speech recognition baseline,'' in \emph{2017 20th Conference of the Oriental Chapter of the International Coordinating Committee on Speech Databases and Speech I/O Systems and Assessment (O-COCOSDA)}, 2017, pp. 1--5.

\bibitem{shi2021aishell3}
Y.~Shi, H.~Bu, X.~Xu, S.~Zhang, and M.~Li, ``{AISHELL-3: A Multi-Speaker Mandarin TTS Corpus},'' in \emph{{Interspeech 2021}}, 2021, pp. 2756--2760.

\bibitem{yamagishi2019cstr}
\BIBentryALTinterwordspacing
J.~Yamagishi, C.~Veaux, and K.~MacDonald, ``Cstr vctk corpus: English multi-speaker corpus for cstr voice cloning toolkit (version 0.92),'' 2019, [sound]. [Online]. Available: \url{https://doi.org/10.7488/ds/2645}
\BIBentrySTDinterwordspacing

\bibitem{zhou2022esd}
\BIBentryALTinterwordspacing
K.~Zhou, B.~Sisman, R.~Liu, and H.~Li, ``Emotional voice conversion: Theory, databases and esd,'' \emph{Speech Communication}, vol. 137, pp. 1--18, 2022. [Online]. Available: \url{https://www.sciencedirect.com/science/article/pii/S0167639321001308}
\BIBentrySTDinterwordspacing

\bibitem{cao2014crema}
H.~Cao, D.~G. Cooper, M.~K. Keutmann, R.~C. Gur, A.~Nenkova, and R.~Verma, ``Crema-d: Crowd-sourced emotional multimodal actors dataset,'' \emph{IEEE Transactions on Affective Computing}, vol.~5, no.~4, pp. 377--390, 2014.

\bibitem{duan2013nus}
Z.~Duan, H.~Fang, B.~Li, K.~C. Sim, and Y.~Wang, ``The nus sung and spoken lyrics corpus: A quantitative comparison of singing and speech,'' in \emph{2013 Asia-Pacific Signal and Information Processing Association Annual Summit and Conference}, 2013, pp. 1--9.

\bibitem{takamichi2019jvs}
S.~Takamichi, K.~Mitsui, Y.~Saito, T.~Koriyama, N.~Tanji, and H.~Saruwatari, ``Jvs corpus: free japanese multi-speaker voice corpus,'' \emph{arXiv preprint arXiv:1908.06248}, 2019.

\bibitem{tamaru2020jvsmusic}
H.~Tamaru, S.~Takamichi, N.~Tanji, and H.~Saruwatari, ``{JVS-MuSiC}: {Japanese} multispeaker singing-voice corpus,'' \emph{arXiv preprint arXiv:2001.07044}, 2020.

\bibitem{morise2016world}
M.~Morise, F.~Yokomori, and K.~Ozawa, ``World: A vocoder-based high-quality speech synthesis system for real-time applications,'' \emph{IEICE TRANSACTIONS on Information and Systems}, vol. E99-D, no.~7, pp. 1877--1884, July 2016.

\end{thebibliography}

\end{document}